\documentclass[letterpaper,twocolumn,english,prl,aps,showpacs,superscriptaddress]{revtex4}
\usepackage{ae,aecompl}
\usepackage[T1]{fontenc}
\usepackage[latin9]{inputenc}
\usepackage{babel}
\usepackage{units}
\usepackage{amsmath}
\usepackage{amssymb}
\usepackage{graphicx}
\usepackage[unicode=true,pdfusetitle,
 bookmarks=true,bookmarksnumbered=false,bookmarksopen=false,
 breaklinks=false,pdfborder={0 0 1},backref=false,colorlinks=false]
 {hyperref}
\usepackage{breakurl}
\usepackage{subfigure}

\makeatletter

 \@ifundefined{textcolor}{}
 {%
   \definecolor{BLACK}{gray}{0}
   \definecolor{WHITE}{gray}{1}
   \definecolor{RED}{rgb}{1,0,0}
   \definecolor{GREEN}{rgb}{0,1,0}
   \definecolor{BLUE}{rgb}{0,0,1}
   \definecolor{CYAN}{cmyk}{1,0,0,0}
   \definecolor{MAGENTA}{cmyk}{0,1,0,0}
   \definecolor{YELLOW}{cmyk}{0,0,1,0}
 }

\newcommand{\Zm}{Z$_{\text{max}}$ }
\newcommand{\zm}{Z$_{\text{max}}$}

\newcommand{\Sm}{$s_{\text{max}}$ }
\newcommand{\sm}{$s_{\text{max}}$}

\newcommand{\Et}{Et$_{12}$ }

\newcommand{\Zt}{Z$_{\text{tot}}$ }

\newcommand{\Mev}{MeV/A }
\newcommand{\mev}{MeV/A}

\newcommand{\Xesn}{$^{129}$\text{Xe}+$^{nat}$\text{Sn} }

\newcommand{\etzt}{(Et$_{12}\times$Z$_{\text{tot}}$)}
\newcommand{\Etzt}{(Et$_{12}\times$Z$_{\text{tot}}$) }

\makeatother

\begin{document}

\title{Nuclear multifragmentation time-scale and fluctuations of largest
fragment size}

\author{D. Gruyer}

\author{J.D. Frankland}

\thanks{Corresponding author}

\email{john.frankland@ganil.fr}

\affiliation{GANIL, CEA-DSM/CNRS-IN2P3, Bvd. Henri Becquerel, F-14076 Caen CEDEX,
France}

\author{R. Botet}

\affiliation{Laboratoire de Physique des Solides, CNRS UMR8502, Université de
Paris-Sud bât. 510, F-91405 Orsay CEDEX, France}

\author{M. P\l{}oszajczak}

\author{E. Bonnet}

\author{A. Chbihi}

\affiliation{GANIL, CEA-DSM/CNRS-IN2P3, Bvd. Henri Becquerel, F-14076 Caen CEDEX,
France}

\author{G. Ademard}

\affiliation{Institut de Physique Nucléaire, CNRS/IN2P3, Université Paris-Sud
11, F-91406 Orsay CEDEX, France}

\author{M. Boisjoli}

\affiliation{GANIL, CEA-DSM/CNRS-IN2P3, Bvd. Henri Becquerel, F-14076 Caen CEDEX,
France}

\affiliation{Département de physique, de génie physique et d'optique, Université
Laval, Québec, G1V 0A6 Canada}

\author{B. Borderie}

\affiliation{Institut de Physique Nucléaire, CNRS/IN2P3, Université Paris-Sud
11, F-91406 Orsay CEDEX, France}

\author{R. Bougault}

\affiliation{LPC, CNRS/IN2P3, Ensicaen, Université de Caen, F-14050 Caen CEDEX,
France}

\author{D. Guinet}

\author{P. Lautesse}

\affiliation{Institut de Physique Nucléaire, Université Claude Bernard Lyon 1,
CNRS/IN2P3, F-69622 Villeurbanne CEDEX, France }

\author{L. Manduci}

\affiliation{École des Applications Militaires de l'Énergie Atomique, B.P. 19,
F-50115 Cherbourg, France}

\author{N. Le Neindre}

\affiliation{LPC, CNRS/IN2P3, Ensicaen, Université de Caen, F-14050 Caen CEDEX,
France}

\author{P. Marini}

\affiliation{GANIL, CEA-DSM/CNRS-IN2P3, Bvd. Henri Becquerel, F-14076 Caen CEDEX,
France}

\author{P. Paw\l{}owski}

\affiliation{H. Niewodnicza\'{n}ski Institute of Nuclear Physics, PL-31342 Kraków,
Poland}

\author{M. F. Rivet}

\affiliation{Institut de Physique Nucléaire, CNRS/IN2P3, Université Paris-Sud
11, F-91406 Orsay CEDEX, France}

\author{E. Rosato}

\author{G. Spadaccini}

\author{M. Vigilante}

\affiliation{Dipartimento di Scienze Fisiche e Sezione INFN, Universita di Napoli
\textquoteleft{}\textquoteleft{}Federico II,\textquoteright{}\textquoteright{}
I-80126 Napoli, Italy}

\author{J.P. Wieleczko}

\affiliation{GANIL, CEA-DSM/CNRS-IN2P3, Bvd. Henri Becquerel, F-14076 Caen CEDEX,
France}

\collaboration{INDRA collaboration}

\noaffiliation
\begin{abstract}
Distributions of the largest fragment charge, \zm, in multifragmentation
reactions around the Fermi energy can be decomposed into a sum of
a Gaussian and a Gumbel distribution, whereas at much higher or lower
energies one or the other distribution is asymptotically dominant.
We demonstrate the same generic behavior for the largest cluster size
in critical aggregation models for small systems, in or out of equilibrium,
around the critical point. By analogy with the time-dependent irreversible
aggregation model, we infer that \Zm distributions are characteristic
of the multifragmentation time-scale, which is largely determined
by the onset of radial expansion in this energy range.
\end{abstract}


\maketitle

\paragraph{Introduction}

In central heavy-ion collisions at beam energies of $\sim$20-150
\Mev multiple production of nuclear fragments can be observed, compatible
with the quasi-simultaneous break-up of finite pieces of excited nuclear
matter \citep{Fox93,Pea94,Heu94,MDA95,Kun96,Marie1997Hot,Hau98}.
This so-called ``nuclear multifragmentation'' is a fascinating process
\citep{Borderie2008Nuclear} which has long been associated with a
predicted liquid-gas coexistence region in the nuclear matter phase
diagram at sub-critical temperatures and sub-saturation densities
\citep{Sie83,Jaq84,Glendenning1986Liquidgas}. Statistical \citep{Gross1990Statistical,Gil94,Bauer1995Preequilibrium,Bon95,MDA96,Rad97,Bea01,Toke2009Common}
and dynamical \citep{Gre87,Aic91,Bur92,Wad04,Colonna2010Fragmentation}
aspects have been widely studied, and evidences supporting equally
well either a continuous phase transition of the liquid-gas universality
class \citep{Hir84,Gil94,Bauer1995Preequilibrium,Pochodzalla1995Probing,Elli00,Berkenbusch2001EventbyEvent,Sch01,Sri02},
a discontinuous (``first-order'') transition occurring within the
coexistence region \citep{MDA00,Campi2005Partial,Sch01,Ell02,Moretto2002Negative,Sri02,DAgostino2002Reliability},
or indeed the survival of initial-state correlations in a purely dynamical
picture \citep{QMD1997initialfinalstatecorrelations,Zbiri2007transitionparticipantspectator}
have been presented. This state of affairs well demonstrates the difficulty
of quantitatively identifying a phase transition in small systems
such as atomic nuclei, where finite-size effects blur the nature of
the transition \citep{Gul99,Car02,Behringer2006Continuous} whose
order may indeed change with the size of the system \citep{Sch01,Sri02},
along with the importance of long-range Coulomb forces \citep{Lev85,MDA95,Sri01,Toke2009Common},
and presence of dynamical effects such as radial flow \citep{Buc84,Pei89,Barz1991Flow,Bauer1993ExplosiveFlow,Kun95,Pak96,Das04}.

In this context we have tried to establish generic features of multifragmentation
in order to deduce its nature in a less model-dependent way. In our
previous works \citep{PhysRevLett.86.3514}, we used the model-independent
universal fluctuations theory \citep{UniversalFluctuationsBook} to
determine that the order parameter of nuclear multifragmentation is
the size of the largest fragment of each partition. This means that
we know to which class of generic cluster models nuclear multifragmentation
belongs, answering the question raised in \citep{Aichelin198415}:
it is an aggregation phenomenon (``condensation of vapor''), not
a fragmentation process (``shattering of glass''). Next \citep{Frankland2005Modelindependent},
we studied the order parameter distributions for a wide range of data
and showed that, to a first approximation, they tend towards one of
two laws : the Gaussian distribution of the central limit theorem
at the lowest energies, or the Gumbel distribution \citep{FisherTippet,Gumbel1958Statistics}
of extreme value statistics \citep{Gnedenko} at the highest. The
system-size dependence of the energy of transition from one regime
to the other was mapped out and tentatively associated with the observed
behavior of limiting temperatures for finite nuclei \citep{Nato02}. 

In this letter we will study in more detail the transition from one
regime to the other, using new data on largest fragment distributions
for multifragmentation in central collisions at bombarding energies
intermediate between the two asymptotic regimes. We will show that,
in the transition region, these distributions are better approximated
by an admixture of the two asymptotic distributions with proportions
which evolve with the bombarding energy. We will then compare this
behavior with that seen in two generic models of aggregation, for
finite systems around a critical point.

\begin{figure}[ht]
\centering \includegraphics[clip,width=0.9\columnwidth]{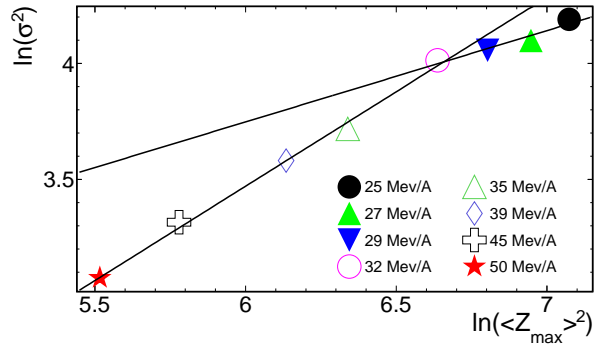} \caption{(color on-line) (Symbols) Log-log correlation between the first two
cumulant moments ($\left\langle \text{{\zm}}\right\rangle ^{2}$and
$\sigma^{2}$) of the order parameter distribution; (Lines) linear
fits performed in the range 50-32 \Mev (resp. 32-25 \mev) which
correspond to a slope $\Delta\sim1$ (resp. $\Delta\sim\nicefrac{1}{2}$)
(see text).\label{fig:deltaScaling} }
\end{figure}

\paragraph{Experimental analysis}

Collisions of \Xesn were measured using the INDRA $4\pi$ charged
product array \citep{Pouthas1995418} at the GANIL accelerator facility.
This detector, composed of 336 detection cells arranged according
to 17 rings centered on the beam axis, covers 90\% of the solid angle
and can identify fragments from hydrogen to uranium with low thresholds.
More details can be found in \citep{Frankland2005Modelindependent,Bonnet2009Fragment}.
Data used here were taken during two separate campaigns: beam energies
of 25, 32, 39, 45 and 50 \Mev were previously analyzed in \citep{PhysRevLett.86.3514,Frankland2005Modelindependent};
measurements at beam energies of 27, 29, and 35 \Mev were subsequently
performed specifically in order to probe the energy range around the
change of scaling regime observed in \citep{PhysRevLett.86.3514},
and are presented here for the first time.

We want to study central collisions, requiring the geometrical overlap
between projectile and target to be as close as possible to total,
while detecting all fragments event by event in order to correctly
measure the distribution of the largest fragment charge, \zm. We
therefore look for events which maximize the quantity \etzt: \Et
is the total transverse energy of light charged particles ($Z=1,2$),
which increases with collision centrality \citep{Plagnol1999Onset},
while \Zt is the sum of the atomic numbers of all detected charged
products in each event. For each beam energy we define a centrality
cut corresponding to the last percentile of the \Etzt distribution
measured with the on-line trigger condition (charged product multiplicity
$M\geq4$).

We begin by examining the scaling properties of \Zm fluctuations
including the three new data points at 27, 29 and 35 \mev. Figure
\ref{fig:deltaScaling} shows the correlations between the first two
cumulant moments of the \Zm distribution ($\left\langle \text{{\zm}}\right\rangle ^{2}$and
$\sigma^{2}$), for each beam energy. As in Fig.1(c) of \citep{PhysRevLett.86.3514},
the data fall on two branches, $\sigma^{2}\sim\left\langle \text{{\zm}}\right\rangle ^{2\Delta}$,
with different values for the scaling parameter, $\Delta$ \citep{Botet2000Universal,UniversalFluctuationsBook}:
$\Delta\sim1$ above 32 \mev, and $\Delta\sim\nicefrac{1}{2}$ below
32 \mev. The new data points present a consistent behavior which
follows the systematic scaling trend. 
\begin{figure}
\subfigure{\includegraphics[clip,width=0.9\columnwidth]{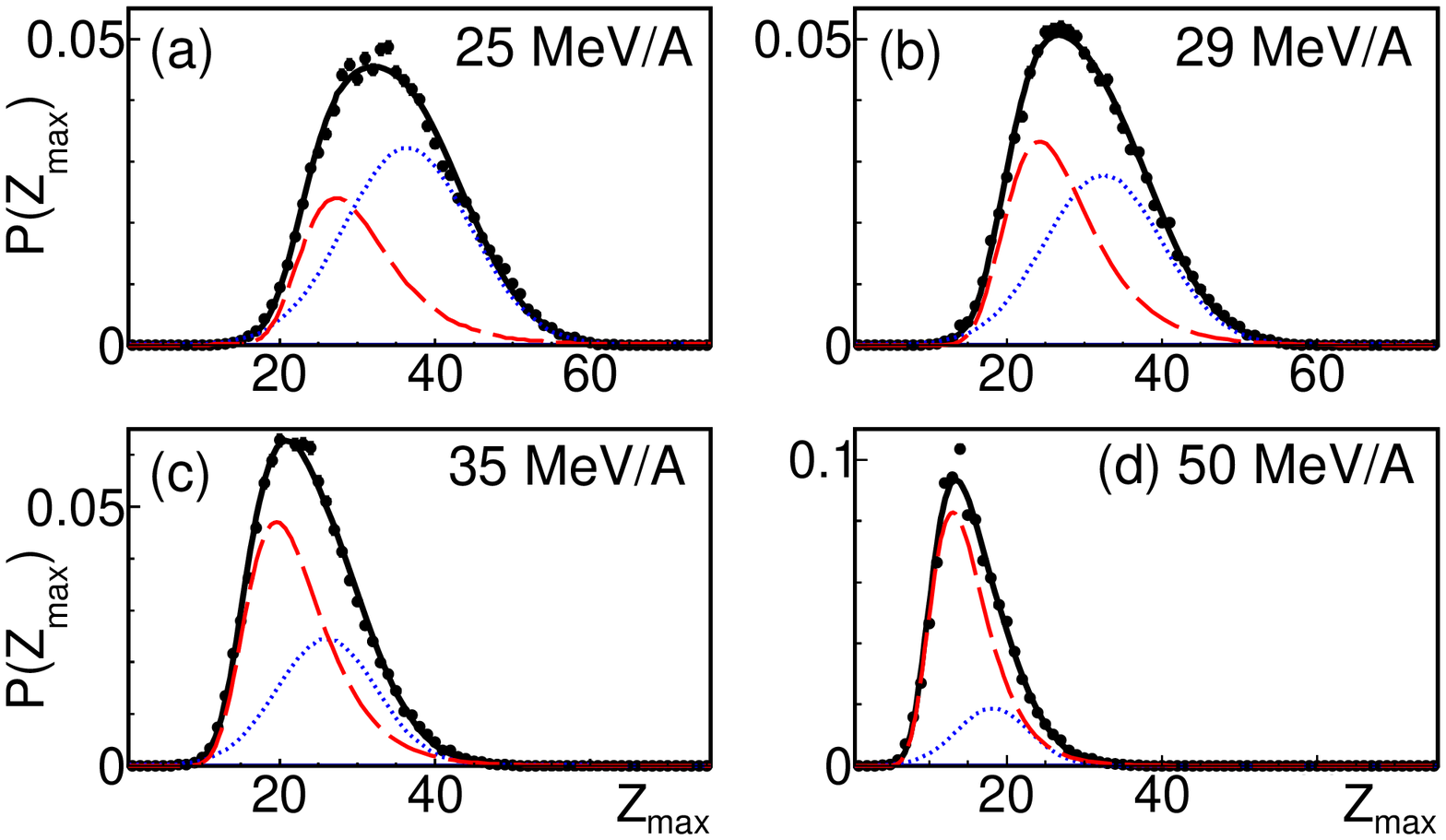}

}

\subfigure{\includegraphics[clip,width=0.8\columnwidth]{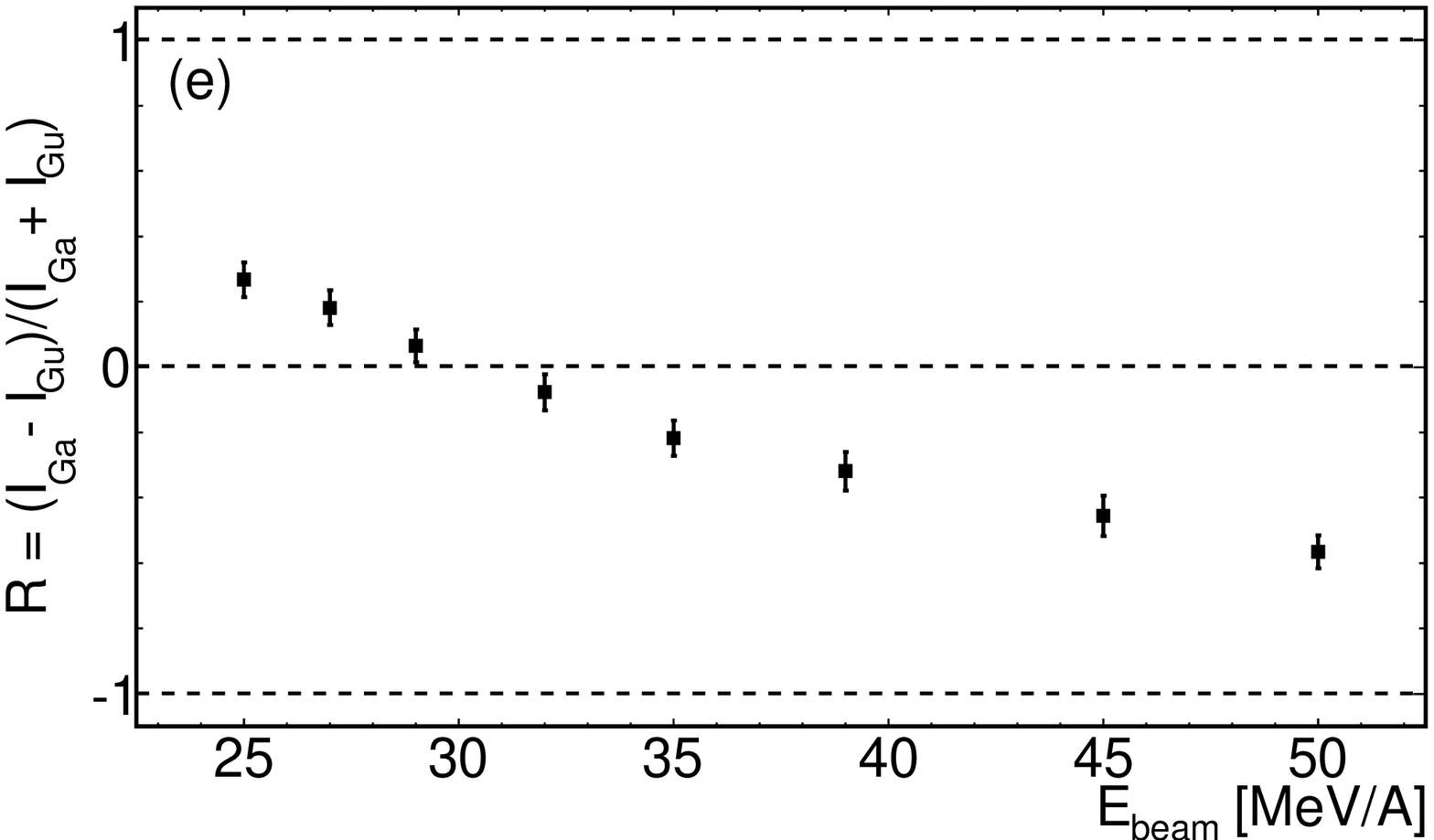}}

\centering{}\caption{(color on-line) (a-d) (dots) \Zm distributions for \Xesn central
collisions for different beam energies; (full line) fit using Eq.(\ref{fitEqua});
(red dashed line) Gumbel component; (blue dotted line) Gaussian component.
(e) Relative strengths of the two components, $R$ (see text), as
a function of the beam energy.\label{fig:fitXeSn} }
\end{figure}

\begin{figure*}[!t]
\subfigure{\includegraphics[clip,height=4.5cm]{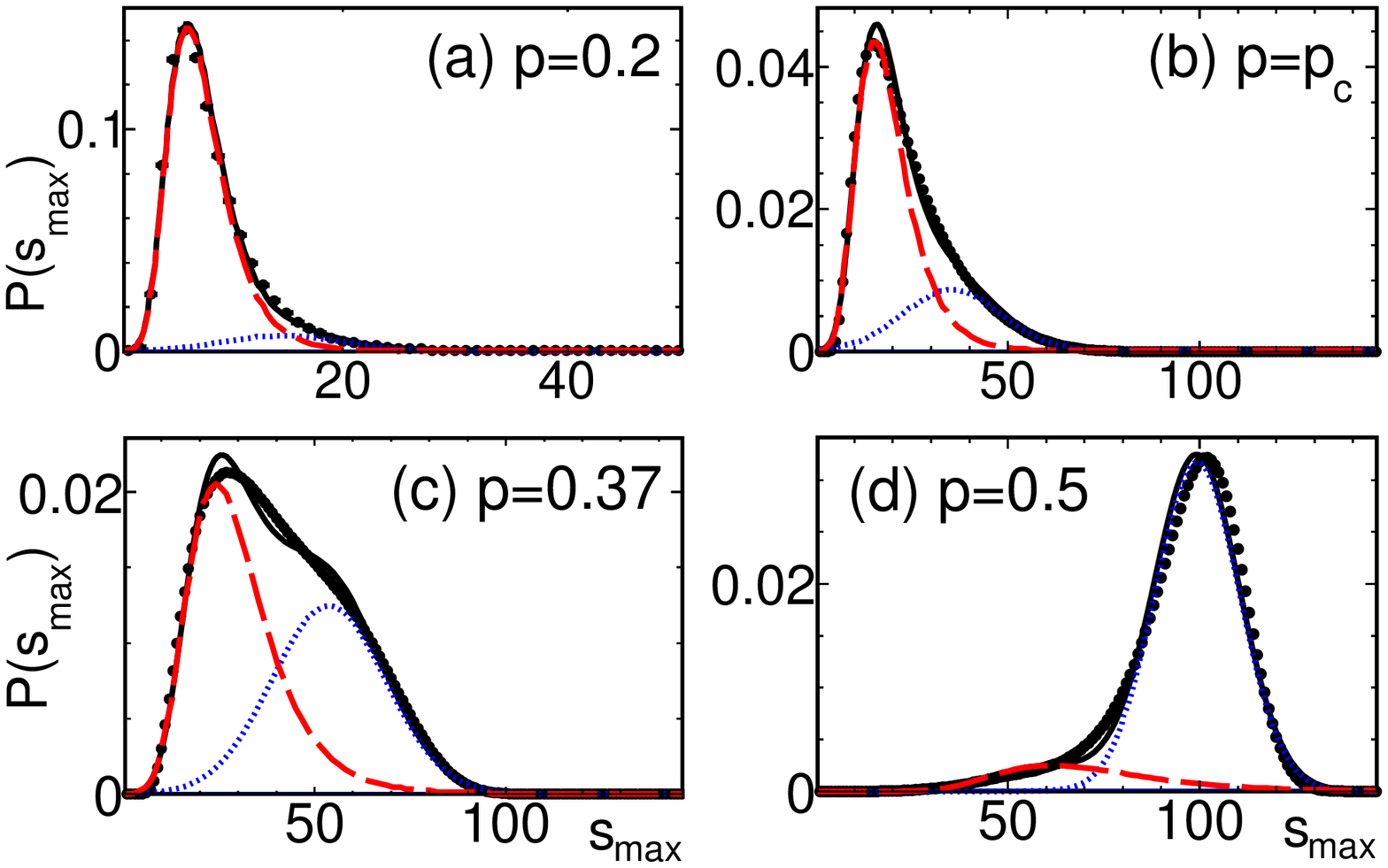}}\hspace*{0.05\paperwidth}\subfigure{\includegraphics[clip,height=4.5cm]{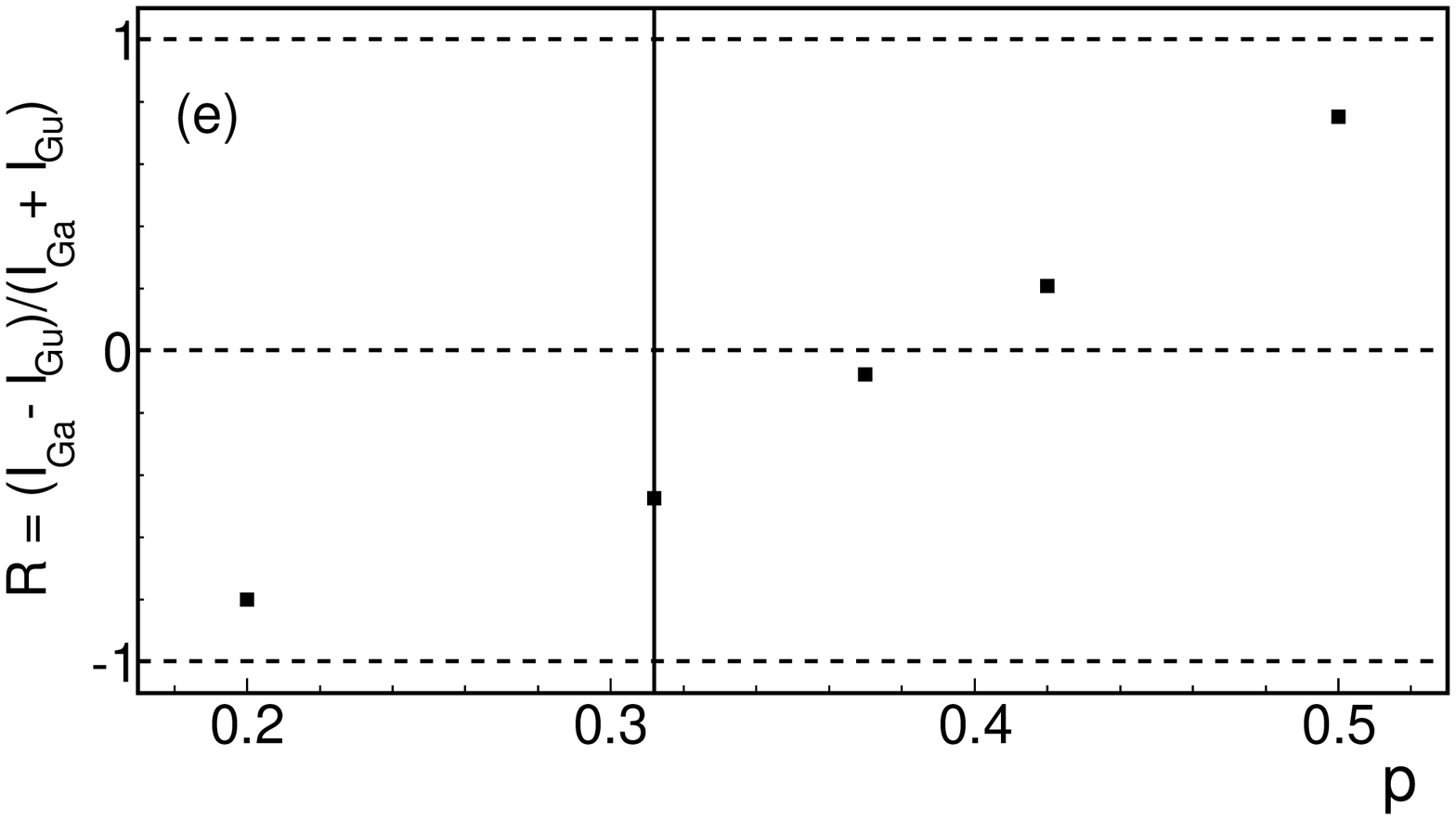}}

\caption{(color online) (a-d) (dots) \Sm distributions for bond percolation
on a $6\times6\times6$ lattice for different values of the bond probability,
$p$; (full line) fit using Eq.(\ref{fitEqua}); (red dashed line)
Gumbel component; (blue dotted line) Gaussian component. (e) Relative
strengths of the two components, $R$ (see text), as a function of
$p$. Vertical line indicates the critical bond probability $p_{c}=0.312$.\label{fig:percolation}}
\end{figure*}
The order parameter distributions (largest fragment charge, \zm,
distributions) for four bombarding energies are presented in Fig.\ref{fig:fitXeSn}(a-d).
They become more asymmetric with increasing bombarding energy, tending
towards the asymptotic Gumbel distribution at 50 \mev. At intermediate
energies (see e.g. new data at 29 \mev, Fig.\ref{fig:fitXeSn}(b))
the distribution is clearly of neither one or the other asymptotic
forms, leading us to propose the following ansatz: the effective \Zm
distribution at any beam energy is an admixture of a Gaussian and
a Gumbel distribution,
\begin{equation}
f(x)=\eta f_{Gauss}(x)+(1-\eta)f_{Gumbel}(x)\label{fitEqua}
\end{equation}
 with $0<\eta<1$ and $x=\text{\zm}$. As a first test of its validity,
fits to the \Zm distribution using Eq.(\ref{fitEqua}) are shown
in Fig.\ref{fig:fitXeSn}. It should be noted that in these fits the
positions and widths of the two components, as well as their relative
weights, were left as free parameters. Reduced $\chi^{2}$ values
for all energies lie between approximately 4 and 10, and can further
be reduced if \Zm distributions are smoothed to remove odd-even staggering
of the yields \citep{Yang1999Oddeven}: in that case all $\chi^{2}$
values are close to 2, except for data at 32 \Mev for which a significantly
larger value $(\chi^{2}\sim4)$ is obtained. These values are significantly
better than those obtained for single-component fits (as in \citep{Frankland2005Modelindependent}).

The quantity $R=2\eta-1$ has been defined in such a way that $R=\pm1$
for pure Gauss/Gumbel distributions, and $R=0$ when both are equally
important. Its evolution over the bombarding energy range 25-50 \Mev
is presented in Fig. \ref{fig:fitXeSn}(e). The vertical bars show
an estimated uncertainty coming from the fitting procedure. The monotonically
decreasing value of $R$ reflects the continuous evolution of the
form of the order parameter distribution. It is interesting to note
that the value $R=0$ is reached between 29 and 32 \mev, in the same
bombarding energy range as the change of $\Delta$-scaling (Fig. \ref{fig:deltaScaling}).

\paragraph{Theoretical models}

We will now consider some results of generic aggregation models in
order to see if our ansatz, Eq.(\ref{fitEqua}), is just a convenient
fitting function or if it can have some physical interpretation. 

First of all, let us consider the most well-known and widely-used
model of this kind, percolation \citep{StaufferPerco}. Indeed, percolation
models have long been used in the analysis and interpretation of multifragmentation
data \citep{Bauer1985New,Campi1988Perco,Ell00,Sisan2001Intermediate,Berkenbusch2001EventbyEvent,Cole02,Brzychczyk2006Largest}.
In bond percolation, each lattice site corresponds to a monomer, and
a proportion $p$ of active bonds is set randomly between sites. Then
clusters of size $s$ are defined as an ensemble of $s$ occupied
sites connected by active bonds. For a definite value of $p=p_{c}$,
a macroscopic cluster appears, corresponding to the sol-gel transition.
The order parameter of the transition is the size of the largest cluster,
\sm, and it is known that for sub-critical finite lattices, \Sm
has a Gumbel distribution \citep{Bazant2000Largest}, while at criticality
in the mean-field limit \Sm follows the Kolmogorov-Smirnov (K-S)
distribution \citep{Botet2005Exact}.

Fig.\ref{fig:percolation}(a-d) show \Sm distributions calculated
for cubic $6\times6\times6$ lattices with bond probabilities $p$
above and below the critical value of $p_{c}=0.312$. For $p\ll p_{c}$
the distribution is well described by the Gumbel distribution alone,
as expected \citep{Bazant2000Largest}, while for $p\gg p_{c}$ the
distribution approaches a Gaussian form. In the critical region the
distribution of \Sm takes on a non-trivial form which is rather well
fitted by the ansatz of Eq.(\ref{fitEqua}). It should be noted that
deviations close to $p=p_{c}$ are to be expected, as in this case
correlations cannot be neglected as required for either Gumbellian
or Gaussian statistics to be strictly valid. Nevertheless the ansatz
of Eq.(\ref{fitEqua}) is a good approximation to the order parameter
distribution for all values of $p$.

Fig.\ref{fig:percolation}(e) presents the evolution of the relative
strengths of the two components, $R$, in the critical domain. It
varies smoothly from $-1$ to $+1$ with increasing $p$. The value
$R=0$ corresponds to very large fluctuations of \sm (see Fig.\ref{fig:percolation}(c)),
but is reached for $p>p_{c}$, not at the infinite-lattice critical
point, due to the finite size of the lattice. It is interesting to
note that the value $R\approx-0.45$ at $p=p_{c}$ is very close to
the value obtained from a fit to the K-S distribution with Eq.(\ref{fitEqua}).
\begin{figure*}[!t]
\subfigure{\includegraphics[clip,height=4.5cm]{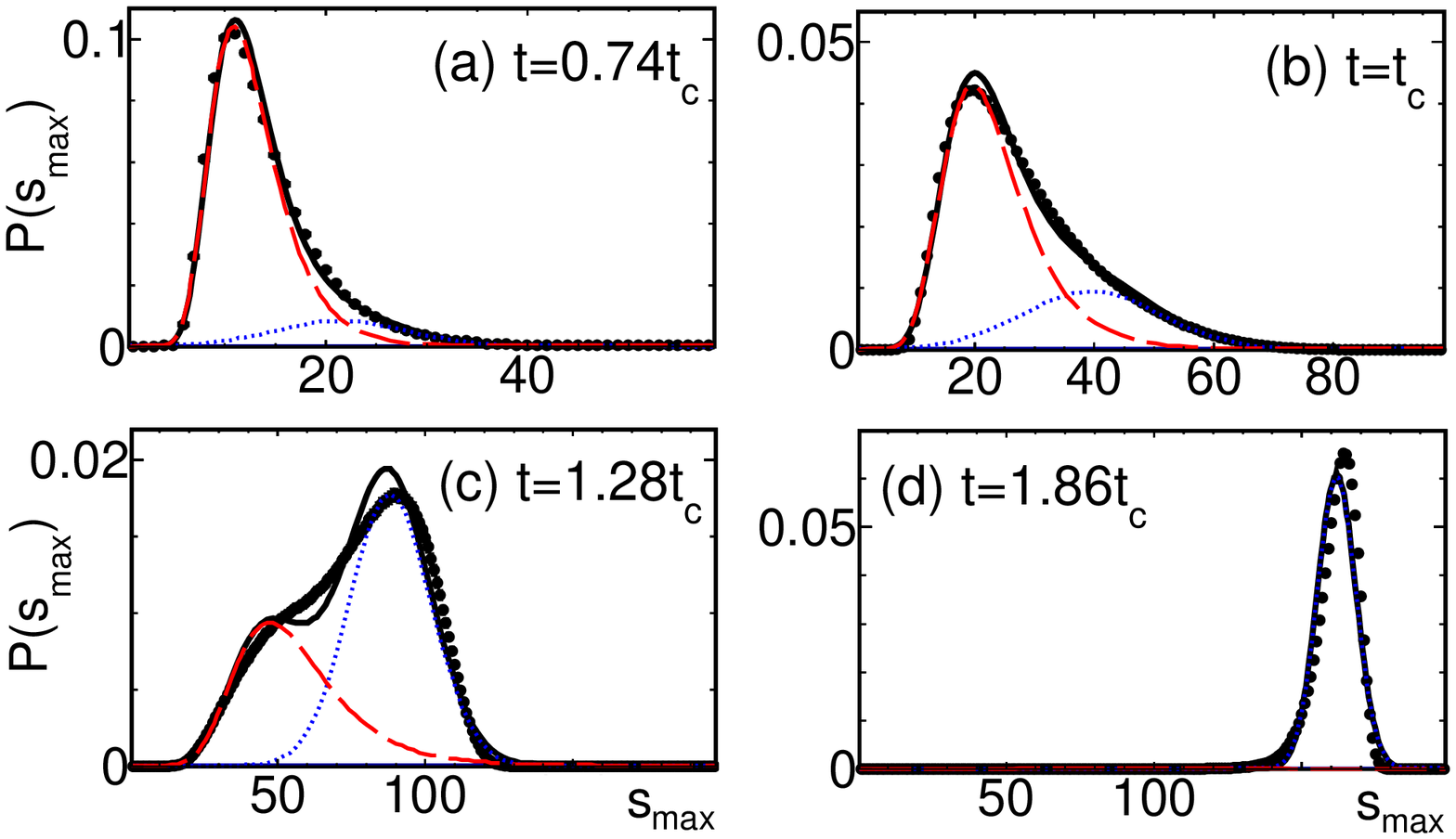}

}\hspace*{0.05\paperwidth}\subfigure{\includegraphics[clip,height=4.5cm]{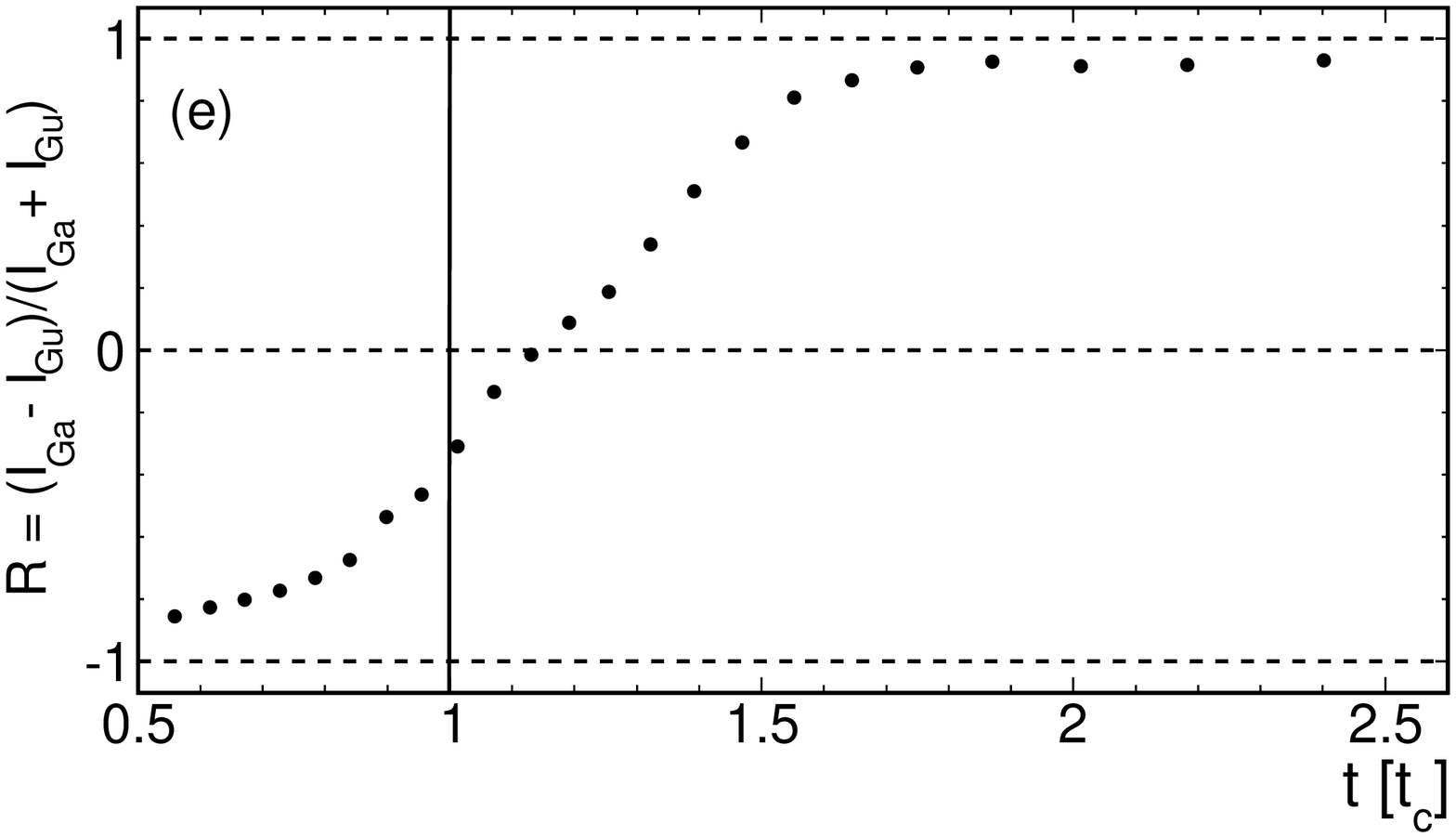}

}

\caption{(color on-line) (a-d) (dots) \Sm distributions for irreversible aggregation
in a system of size $N=216$ for different times; (full line) fit
using Eq.(\ref{fitEqua}); (red dashed line) Gumbel component; (blue
dotted line) Gaussian component. (e) Relative strengths of the two
components, $R$ (see text), as a function of time with respect to
$t_{c}$. Vertical line indicates critical gelation time.\label{fig:smoluchowski}}
\end{figure*}

Percolation is an equilibrium model of reversible aggregation. The
irreversible sol-gel transition can be modeled using the coupled nonlinear
differential equations in the concentrations of clusters of mass $s$
per unit volume, known as the Smoluchowski equations \citep{PhysRevLett.54.1396}.
The probability of aggregation per unit time between clusters is a
homogeneous function of cluster masses, and after a time $t_{c}$,
called the critical gelation time, an ``infinite'' cluster appears.
It was shown in \citep{Botet2000Universal} that for this model \Sm
exhibits the $\Delta=1$ scaling at $t=t_{c}$, and $\Delta=\nicefrac{1}{2}$
scaling with a symmetric distribution for $t\gg t_{c}$. Fig.\ref{fig:smoluchowski}(a-d)
present the \Sm distributions obtained at different times for a system
of total mass $N=216$. These distributions and their evolution over
time are very similar to those obtained from percolation as a function
of the bond probability, Fig.\ref{fig:percolation}. Fits to these
distributions using Eq.(\ref{fitEqua}) are shown, and once again
they are a good approximation for \Sm distributions both far from
and around the critical region. Fig.\ref{fig:smoluchowski}(e) shows
the relative strengths of the two components, $R$, obtained from
the fits as a function of time: as in the case of percolation calculations,
$R=0$ does not coincide with the infinite-system critical gelation
time (in fact it occurs between $t=t_{c}$ and the time of maximum
fluctuations). Also, it may be remarked that at the critical time,
the value of $R$ is very close to the K-S distribution one, and indeed
independent of the size, $N$, of the system. Moreover, these results
are unchanged if a fragmentation kernel is included in the equations,
\emph{i.e. }if clusters can also split into smaller pieces \citep{Botet2011WPCF}.

\paragraph{Discussion}

We have shown that there is a strong similarity between \Zm distributions
for central collisions of \Xesn and the order parameter distributions
obtained for two very different generic aggregation models in their
critical domain. We would now like to understand this similarity.
The percolation model offers little scope for interpretation, as all
physics is ``hidden'' in the bond probability, $p$. On the other
hand, the physical picture of clusters being built up over time by
agglomeration described by the Smoluchowski equations recalls microscopic
approaches in which fragments result from the spinodal decomposition
of the hot, expanding nuclear matter formed by the head-on collision
of two nuclei \citep{Ber83,Lope89,Cho04,Tabacaru2003Fragment}.

In this framework, we can infer that the \Zm distribution and its
form, quantified by the ratio $R$, reflect the time-scale of fragment
formation, whose determining factor is the amount of collective radial
expansion which increases with bombarding energy \citep{Borderie2008Nuclear}.
It has been shown that for central \Xesn reactions the onset of significant
radial expansion occurs for beam energies above $25$ \Mev \citep{Bonnet2009Fragment}.
The similarity between Figs. \ref{fig:fitXeSn} and \ref{fig:smoluchowski}
can therefore be understood in terms of fragment size distributions
being determined on shorter and shorter time-scales due to increasingly
rapid expansion.

We can use this interpretation to understand the system mass-dependence
of the energy of transition from $\Delta=\nicefrac{1}{2}$ to $\Delta=1$
scaling presented in \citep{Frankland2005Modelindependent}: for $^{58}$Ni+$^{58}$Ni
collisions, measured from 32 \Mev to 90 \mev, a change of $\Delta$-scaling
and of the form of the \Zm distribution were observed, as for \Xesn
but at a higher bombarding energy of 52 \mev; for the lighter system
$^{36}$Ar+KCl, studied from 32 \Mev to 74 \mev, only the $\Delta=\nicefrac{1}{2}$
regime was observed, with quasi-Gaussian \Zm distributions; on the
other hand, for the much heavier $^{197}$Au+$^{197}$Au system, at
bombarding energies between 40 \Mev and 80 \mev, only the $\Delta=1$
regime occurs, with Gumbel \Zm distributions. 

Radial expansion in central heavy-ion collisions occurs after significant
compression of the incoming nuclear fluid, and as such depends not
only on static nuclear matter properties such as incompressibility,
but also on transport properties such as the degree of stopping achieved
in the collision \citep{Reis04}. The latter increases with the mass
of the colliding nuclei, as shown in \citep{Lehaut2010Study}. Thus
for light systems, such as $^{36}$Ar+KCl or $^{58}$Ni+$^{58}$Ni,
the bombarding energy required to achieve sufficient initial compression
for there to be significant radial expansion is higher than for the
heavier systems like \Xesn and $^{197}$Au+$^{197}$Au. This explains
why the $\Delta$-scaling transition occurs at higher energy for $^{58}$Ni+$^{58}$Ni
than for \Xesn. For the very light $^{36}$Ar+KCl system the threshold
is higher still than for $^{58}$Ni+$^{58}$Ni, outside the range
of measured bombarding energies. On the other hand, for $^{197}$Au+$^{197}$Au
both the greater initial compression and the far larger Coulomb contribution
may come into play in order to reduce the fragment formation time-scale
even at the lowest bombarding energy.

\paragraph{Conclusions}

We have shown that, for finite systems, the largest cluster size distribution
in critical aggregation models is an admixture of the two asymptotic
distributions observed far below and above the critical region. This
result holds true for both equilibrium (percolation) and out-of-equilibrium
(irreversible aggregation) models. A similar decomposition has been
shown for the experimentally-observed charge distribution of the largest
fragment per event produced in nuclear multifragmentation, indicating
that the critical domain lies around $E_{beam}\approx30$\Mev for
the \Xesn system. By analogy with the irreversible aggregation model,
where the form of the order parameter distribution depends on the
time-scale of the process, we interpret such criticality along with
the corresponding change of $\Delta$-scaling as the onset of an ``explosive''
multifragmentation regime in which initially compressed heated nuclear
liquid clusterizes in the presence of significant radial expansion
\citep{Bauer1993ExplosiveFlow,Rei04}. The mass-dependence of the
energy at which the onset occurs for different (symmetric) systems
is related to nuclear stopping and hence to transport properties of
hot nuclear matter. Such an overall picture is both consistent with,
and provides a link between, recent results on the role of radial
expansion in nuclear multifragmentation \citep{Bonnet2009Fragment,Bonnet2010New}
and the systematic study of nuclear stopping around the Fermi energy
\citep{Lehaut2010Study}.
\begin{acknowledgments}
The authors would like to thank the staff of the GANIL Accelerator
facility for their continued support during the experiments. D. G.
gratefully acknowledges the financial support of the Commissariat
à l'Énergie Atomique and the Conseil Régional de Basse-Normandie.
This work was also supported by the Centre National de la Recherche
Scientifique and by the Ministère de l'Éducation Nationale.
\end{acknowledgments}
\bibliographystyle{apsrev}
\bibliography{paper}

\end{document}